\documentclass{elsart}
\usepackage{natbib}

\newcommand{\la}{\langle}
\newcommand{\ra}{\rangle}
\newcommand{\lla}{\left\langle}
\newcommand{\rra}{\right\rangle}
\newcommand{\be}{\begin{eqnarray}}
\newcommand{\ee}{\end{eqnarray}}
\newcommand{\ben}{\begin{eqnarray*}}
\newcommand{\een}{\end{eqnarray*}}
\def\lab#1	{\hbox{\small #1} }
\def\mb#1       {\mbox{\boldmath $#1$}}

\def\mb#1         {\mbox{\boldmath $#1$}}
\def\diffn#1	  {\Delta^{-}_{#1}}

\def\la           {\left\langle }
\def\ra           {\right\rangle }

\begin{document}
\begin{frontmatter}

\begin{tabbing}
\` {\sl hep-lat/9809094} \\
    \\
\` LSUHE No. 270-1998 \\
\` August, 1998 \\
\end{tabbing}
 
\vspace*{0.1in}

\title{Confinement studies in lattice QCD\thanksref{X}}
\author[haymaker]{Richard W. Haymaker}
\thanks[X] {Talk given at the Symposium in Honor of
Dick Slansky, Los Alamos, May 20-21 1998}
\address[haymaker]{Department of Physics and Astronomy, 
Louisiana State University,  Baton Rouge, Louisiana 70803 USA}
\begin{abstract}
We describe the current search for confinement mechanisms 
in lattice QCD.  We report on a recent  derivation  of a lattice
Ehrenfest-Maxwell relation  for the Abelian projection of SU(2)
lattice gauge theory.   This gives a precise lattice definition of 
field strength and  electric current due to static sources,
charged dynamical fields,  gauge fixing and ghosts. 
In the maximal Abelian gauge the electric charge is 
anti-screened analogously to the non-Abelian charge.
\end{abstract}
\end{frontmatter}

\section{Introduction}

Quenched lattice QCD calculations of the static quark anti-quark
potential have firmly established a linearly rising behavior 
over all distances obtainable in state-of-the-art simulations\cite{bss1}.
However this situation satisfies no one, least of all a Dick 
Slansky.   A `black box' calculation at limited
quark separation gives supporting evidence that QCD 
confines quarks, but it offers no explanation nor reveals 
a principle governing the phenomenon.

Lattice QCD is more than just an algorithm to calculate  quantities 
at strong coupling.   The lattice is a regulator for QCD, 
parametrized by the lattice spacing $a$. Gauge symmetry is preserved at
all costs.  This regulation scheme is perhaps the only one that 
gives a completely self consistent cutoff model in its own right.  
The dynamical variables are elements of the symmetry group rather than
the Lie algebra.  As a consequence, 
 many of the topological features that are  `likely suspects' 
in the physics of confinement have  natural definitions
on the lattice.   These include conserved  U(1), 
Z(N), SU(N)/Z(N) monopole loops, 
Dirac sheets,  Z(N) and SU(N)  vortex sheets 
and other features.

Lattice work\cite{creutzbook,montveymunster,rothe,kogut,latxx} 
is coming into its own in sorting through some of the
seminal ideas on confinement\cite{no,ks,nambu,parisi,mandelstam,thooft4}
 proposed in the '70's.  I venture to
say that Dick Slansky could easily have been drawn into the fray because it
has all the elements that typically captured his imagination.   The 
central questions are fundamental.  
They involve interesting issues in group theory,
topology, and duality.   Candidate mechanisms are proliferting and 
fundamental questions remain unsettled\footnote{At Lattice '98, the XVI International
Symposium on Lattice Field Theory in Boulder, there were 52
contributed papers on color confinement, up from a dozen or so in the
early '90's.}.

Confinement is a consequence of a disordered state characterized by
the expectation value of a Wilson loop suppressed to an area 
law rather than a perimeter law for an ordered system.  
Contributions from  topological objects e.g. monopoles and 
vortices  can accomplish this.  On the lattice these objects are
not singular.  They occur abundantly in SU(N) lattice theories.
 It is only in the continuum limit of zero 
lattice spacing where these approach singular structures.  In lattice
simulations, one can tamper with these objects and see if suppression
is accompanied by the loss of an area law.   In this way one
has a laboratory to study candidate disordering mechanisms.  

 It is not realistic to try to give a proper review of the many 
 competing views
in this brief article.  Further,  the studies are possibly diverging
more rapidly than at any point in the past considering  the variety of
papers on the subject over the past year.   However I would like to
touch on them and further to 
argue that all the studies are describing genuine properties of 
QCD, seen through the eyes  of subsets of the full dynamical variables.
The controversies have to do with which scenario will lead to
the most compelling  explanation of confinement.

Although these lattice studies reveal relevant confinement physics, 
the goal is not completely clear.  Most would agree however
that if we knew the dual form of QCD  it would give a 
definitive description of the physics of confinement.

Consider the example of superconductivity.  By  identifying 
the carriers of 
the persistent current one can discover an instability in the normal
vacuum.   The Ginzburg-Landau effective theory describes the consequence
of the instability which of course elucidates  the fundamental
principle  underlying the phenomenon which is the spontaneous
breaking of the U(1) gauge symmetry\cite{weinberg}.
We might imagine that an understanding of the topological
structures might in the same way lead to a discovery of 
a spontaneous gauge symmetry breaking.  The problem is
that a definitive scenario still seems quite distant.

\section{Lattice  confinement studies - a brief survey}

In this section I wish to call attention  
to some  of the approaches in lattice studies 
of confinement that have 
been active over the past 12 months.  This is principally
to give a flavor of the work and a list of recent references,
and rely in the references therin for a more complete bibliography.

Our most recent work, which we
describe in Sec. 5 is based on the Abelian projection. 

We will restrict our attention to
SU(2) theories throughout this article,  believing that 
the essential issues will
be revealed in this simplest case.

\subsection{Abelian projection, Abelian dominance}

In the Georgi-Glashow model with a gauge field 
coupled to  the adjoint Higgs one can define a gauge
invariant Abelian field strength\cite{thooft2}.  This is the 
construction used to identify the magnetic field
of a 'tHooft-Polyakov monopole.  Further, the Higgs field can
be used to define an Abelian projected theory.
This is accomplished by gauge transforming the Higgs field
into the $3$ direction. The resulting partially gauge fixed theory is
still invariant under U(1) gauge transformations -- rotations
about the $3$ axis.

Pure gauge SU(2) has no adjoint Higgs field and so there
is no straightforward way to define Abelian variables.  It is still
possible to define an Abelian Projection\cite{thooft1,klsw,suzuki}.

One can find a  ``collective" Higgs field transforming 
under the adjoint representation which we denote as an "adjoint field".
Consider an arbitrary Wilson line starting and ending at 
a particular site.   This can be parametrized by
$W(x) = \cos \chi + i  \mb{\phi} \cdot \mb{\sigma} \sin \chi$ where 
$ \mb{\phi} $ is a normalized adjoint field.  The Wilson line
could be (i) an open (no trace) Poliakov loop,  (ii) 
an open plaquette in e.g. the $(1,2)$ plane, or  
(iii) an arbitrary sum of such lines then normalized
to construct an SU(2) element.   One can  generalize to
 (iv) define the adjoint  field  self consistently by
requiring that the adjoint field at one site be equal to  the
normalized sum of the eight neighboring adjoint fields 
after parallel transport to that site.  

In these four examples, one can further fix the gauge by rotating
the adjoint field into the third direction.    This then 
fixes (i)  to the Polyakov gauge, (ii) to a plaquette gauge  (iv) 
e.g. a clover gauge, and (iv) in the maximal Abelian gauge.  
The "collective" adjoint field itself is gauge covariant,
it does not necessitate a gauge fixing. Different adjoint fields
correspond to different Abelian projections.

The maximal Abelian gauge appears to be the most promising choice. 
The condition, equivalent to the one given just above\cite{bdh}, 
 maximizes $\sum (U_4^2 + U_3^2 - U_2^2 - U_1^2)$,
where $ U = U_4 + i \mb{\sigma} \cdot \mb{U} $, 
$U_4^2 + \mb{U} ^2 = 1$.
The maximization brings the links as close as possible to the 
values $U_1 = U_2 = 0$ and $U_4^2 + U_3^2 = 1$, leaving a U(1) 
invariance.

A new and interesting alternative has been proposed 
recently  by Van der Sijs\cite{sijs}.
This gauge fixing is based on the lowest eigenstate of the covariant
Laplacian operator:  i.e. sum the adjoint fields at the eight
neighboring sites - parallel transported to a given site - and subtract
$ 8 \times $ the field at the site.  This defines a gauge with no lattice
Gribov copies.  And it identifies the positions of 
monopoles as singularities (zeros) of  the  adjoint field.

`Abelian dominance' asserts further that operators built out of the
U(1) links defined by the `Abelian projection' will dominate
the  calculation of string tension and other observables.  There are
a large number of numerical tests which support Abelian dominance.
However the successful tests are for the most part restricted to 
(a)  certain specific quantites  e.g. string tension in the fundamental
representation,  and (b) for the gauge choice of the maximal
Abelian gauge.   In other gauges, e.g. plaquette gauge, and/or 
for other quantities such as string tension in other representations 
Abelian dominance is not well supported and in some cases strongly
violated.

On the other hand in a series of papers 
Di Giacomo et. al.\cite{giacomo,gp,baeg,glmp,ddpp}
 have studied the finite
temperature confining phase transition and shown that 
the  disorder parameter is very insensitive to the choice
of Abelian projection.  This is the foundation of the
dual superconductor scenario. 
 This is not in conflict with the above
results because it does not require the Abelian dominance ansatz.

Using the Abelian projection and Abelian dominance,  the 
problem is reduced to a  U(1) gauge theory albeit 
with a complicated action coming from the effects of
gauge fixing.   As in a pure U(1) gauge theory,
Dirac monopoles can be identified 
as the charge carriers of the persistent currents
in the dual superconductor scenario\cite{dt,suzuki,shb,ss,sw}.

Bakker, Chernodub and Polikarpov\cite{bcp} have shown that Abelian 
monopole currents,  defined in the maximal Abelian gauge
are physical objects:  there is a strong local correlation
between monopoles and enhancements in the SU(2) gauge invariant
action.   

 Other recent results of Polikarpov et al\cite{cgp,cgpv,bcgpv} include:
Abelian monopoles in the maximal Abelian gauge are dyons.
There are strong correlations between  magnetic charge,
electric charge and topological charge density.  These
same connections have been reported independently by
Ilgenfritz et al\cite{immt}.

In order to understand the disordering effect of monopoles
on large Wilson loops, 
Hart and Teper\cite{hartteper} have studied the clustering properties of 
monopole loops and found two classes of clusters:  A single
cluster permeates the whole lattice volume, the remaining are
small localized clusters.

In a series of papers,  Ichie and 
Suganuma\cite{is,sita,ts,its} have sought a deeper
understanding of Abelian dominance.  They look at the residual
degrees of freedom after Abelian projection,  i.e. the coset
fields and argue that in a random variable approximation that 
Abelian dominance is exact.

Ogilvie\cite{ogilvie} has argued in the context of the Abelian projection
that gauge fixing  is in principle  unnecessary, that results are 
the same whether the gauge is fixed or not.  This is at odds with
many simulations, leaving much to be sorted out.

Grady\cite{grady1,grady2}  has given evidence that casts doubt on 
whether Abelian monopoles confinement mechanism carries
over to the continuum limit.

A number of authors, including our group  
have reported  that a well defined electric vortex forms between
static sources\cite{sbh,bdh,bali98}.   
The vortex is very well described by a dual
Ginzburg-Landau effective theory.   In effect we are able to show
that there is a local relation between the electric flux and
the curl of the monopole current, defined by Abelian projection
of full SU(2) field configurations that gives a damping of 
fields as they penetrate the dual superconductor.  This directly
accounts for confinement.   In Sec. 5 we derive a precise definition
of the electric field which tightens up the 
definition of vortices\cite{dhh}.

\subsection{Vortices in Z(N) $\times$ SU(N)/Z(N) formulation }

In the early 80's there was an effort by 
Mack and Petkova\cite{mp}, Yaffe\cite{yaffe}, Tomboulis
\cite{tomboulis2}, Yoneya\cite{yoneya},  Cornwall\cite{cornwall}, and
Halliday and Schwimmer\cite{hs}, to understand confinement in terms 
of SU(N) vortices.  These are singular gauge configurations
characterized by their topological properties.  They are
multiple valued in SU(N) but single valued in SU(N)/Z(N).

There are  more recent developments by 
Tomboulis and Kovacs\cite{kt1,kt2,tomboulis1}
\footnote{A thorough review  of the
topological issues in this approach can be found in Ref.\cite{kt1}}.   
Their study is based on a 
decomposition of  the SU(2) partition function  into variables
defined over the center Z(2) and variables defined over 
the quotient group SU(2)/Z(2) (= SO(3))\cite{tomboulis2}. 
Whereas the SU(2) manifold has a trivial topology, SO(3) does not:
$\pi_1$(SO(3)) = Z(2).

To accomplish this reformulation, one appends new Z(2) valued variables 
$\sigma_{p}$, which live on plaquettes, to the usual set of
SU(2) links.   This allows one to write the 
standard SU(2) Wilson action partition function 
in terms of these Z(2) variables, $\sigma_{p}$ and 
link variables, $\{\hat{U}_l\}$, defined 
over the manifold of SO(3) rather than  SU(2).  
The latter distinction is manifest in the reformulated partition
function in that the SU(2) links $\{U_l\}$ occur only in Z(2) invariant
combinations i.e. invariant 
 under the transformation $ U_l \rightarrow \pm U_l $.

Consider a single $(1,2)$ plaquette with $\sigma_p = -1$ , 
all others $= +1$.  A Z(2) monopole occurs in each of the two (1,2,3) 
cubes adjacent to the face of the 
negative plaquette and similarly in the two
(1,2,4) cubes.  
 
On the dual lattice, each cube becomes a dual link 
orthogonal to the cube and this forms 
a conserved Z(2) monopole
current loop (the smallest such loop) 
on the dual lattice.     The negative
plaquettes become  dual plaquettes on the dual lattice.  They
form a surface bounded by the  monopole loop.  This surface is
called a "thin vortex" sheet.  
 
One can build larger monopole  current loops and vortices 
by laying out stacks of negative $\sigma$ plaquettes.
 
There are also the usual plaquettes built from the trace of 4 SU(2)
link variables.  Again consider a  configuration in which all 
such plaquettes
are positive except for a single $(1,2)$  plaquette which is negative.
An SO(3) monopole occurs in each of the two (1,2,3) 
cubes adjacent to the negative plaquette and similarly in the two
(1,2,4) cubes.  This is an SO(3)  object because every link 
involved in its construction occurs quadratically and hence is
invariant under a Z(2) transformation of each link.

Analogously there is an SO(3) monopole current loop on the dual lattice.
The  surface bounded by the loop is a 
a Dirac sheet which differs from the "thin
vortex" sheet in that it can be moved arbitrarily by multiplying
the SU(2) links by elements of the center $\pm$ without costing any
action.

The equivalence of the two forms of the partition function
requires that every Z(2) monopole 
be coincident with an SO(3) monopole and vice versa.  
Hence we can visualize the
vortex structure as two currents on coincident loops,  
bounded by two surfaces. This is called a `hybrid' vortex.  
We can also shrink the monopole loop to zero, 
leaving either a pure "thin vortex"
sheet, or in the other case a pure Dirac sheet.  

The SO(3) monopoles also have associated non-Abelian 
(non-integer valued) flux which costs action.  These configurations
are  denoted "thick vortices",    The Dirac sheet is the
return flux required by the topological group $\pi_1(SO(3)) = Z(2)$.

In this picture, it has been shown that thin vortices are exponentially
suppressed at weak coupling and can not account for confinement.  
However SO(3) thick vortices can occur at weak coupling in which
all terms in the action $|tr U_p| \approx 1$.  In 
recent papers Kovacs and Tomboulis\cite{kt1,kt2} 
have shown that the static quark potential can be reproduced by
contributions from SO(3) thick vortices linking 
the Wilson loop in SU(2); and in SU(3); and 
that exclusion of vortices results in a perimeter law.

See Gavai and Mathur\cite{gm} for a study of Z(2) monopoles
and the deconfinement phase transition.   Also see Grady\cite{grady} for
variation on the SO(3) - Z(2)  monopole construction.

\subsection{Center Projection Vortices}

There are other closely related approaches that also focus on the center
of the group as the key to understanding confinement. 
In a number of recent papers, 
Greensite, Del Debbio, Faber and Olejnik\cite{dfggo} have introduced 
the `Center projection' as a way of identifying Z(2) vortices for SU(2).
which they denote as "Projection vortices".  They differ from 
the thick and thin vortices of Tomboulis.  They are defined in
the the maximal center gauge.   In this gauge, one maximizes
$\sum U_4^2 $.  Hence $U_4$ will be as close as possible to
$\pm 1$.  The Z(2) links  are then defined $Z = sign(U_4) $.  
This gives a Z(2) gauge theory that has "P vortex" excitations
which are somewhat analogous to the Tomboulis `thin' vortices.
They also define "center vortices" which can be 
much larger than one lattice spacing and are analogous
to the Tomboulis "thick vortices".

Greensite et. al.\cite{dfgo,dfgo1,fgo} have reported a growing 
list of encouraging results. Among the results are:
P vortices are well correlated with SU(N) vortices; No P vortices
no confinement; P vortices account for the full string tension; 
P vortex density scales;  Center vortices thicken as the lattice cools;
P-vortex locations are correlated among Gribov Copies;  Preliminary
successful generalizations to SU(3);  and center vortices are compatible 
with Casimir scaling.

See also Langfeld et al\cite{lrt,lter} and  
Stephenson\cite{stephenson} for further
results on center vortices.

A  variation in the center projection procedure is to first
fix to the maximal Abelian gauge, and then maximize $\sum \cos \chi$ 
where $\chi$ is the U(1) link angle.  This is denoted the indirect
maximal center gauge, as opposed to the above procedure 
denoted  the direct maximal
center gauge.   In this gauge, a sheet consisting of monopole loops
alternating with anti-monopole loops coincides with the P vortex 
sheet\cite{dfgo,fgo} indicating a possible 
overlap between this and  other scenarios.

\subsection{Dual variables}

One expects that the disorder
regime in QCD would be described by an ordered regime 
in dual QCD. A definitive form of dual QCD would probably take
us a long way toward an understanding of confinement.

We would like to call attention to recent lattice 
work on dual variables
by Cheluvaraja\cite{cheluvaraja}, and continuum work by
Majumdar and Sharatchandra\cite{ms},
Sharatchandra et al\cite{sharatchandra}, 
Mathur\cite{mathur}, Faddeev and Niemi\cite{fn},  
and Chan and Tsun\cite{ct}. 

The work by Majumdar and Sharatchandra
supports the dual QCD ansatz by Baker, Ball and Zachariasen\cite{bbz}.

Seiberg and Witten\cite{seibergwitten} exploited duality 
to establish confinement for supersymmetric QCD.

\subsection{Instantons}

An important aspect of confinement studies is to identify what objects
can not account for confinement.  Instantons were likely suspects at
one time.     Smoothing techniques are very important
in the identification of instantons.
Unlike monopoles,  small  instantons can fall through the holes in
the lattice and further can be swamped by short distance 
fluctuations.   New smoothing techniques which overcome these
difficulties, denoted `renormalization
group mapping', have
been applied to configurations by 
DeGrand, Hasenfratz and Kovacs\cite{dhk}
 to  elucidate the role of instantons in the QCD vacuum.  The come to 
 the strong conclusion that instantons alone do not confine.  

See also other recent works by Narayan and Neuberger\cite{nn};
Narayan and Vranas\cite{nv};  
Ph. de Forcrand, M. Garcia Perez and I.-O. Stamatescu\cite{fgs}; and
B. Alles, M. D'Elia and A. Di Giacomo\cite{aeg}.

\subsection{Interconnections}

Even though the possible explanations for confinement seem
to be diverging at this point,  I would like to re-emphasize 
that different descriptions can in some
cases be describing the same underlying physics:   

Abelian monopoles in the maximal Abelian gauge correlate with
\begin{itemize}

\item
 SU(2) 
enhancements in the gauge invariant action\cite{bcp},  

\item
gauge invariant topological
charge\cite{cgp,cgpv,bcgpv,immt},

\item
P vortices in the maximal center gauge\cite{dfgo1},

\item
Instantons\cite{cg,ht}.

\end{itemize}

Therefore the monopoles can not be thought of as 
merely gauge artifacts.

\section{U(1) gauge theories and superconductivity}

The onset of superconductivity is governed by the 
spontaneous breaking of the U(1) gauge symmetry via a 
non-zero vacuum expectation value of a 
charged field \cite{weinberg}.    An immediate
consequence of this is the generation of a photon mass and, for type II
superconductors, the formation of magnetic 
vortices which confine magnetic
flux to narrow tubes\cite{hay}
as revealed by the Ginzburg-Landau effective
theory. 
Lattice studies of dual superconductivity in SU(N) gauge theories seek
to exploit this connection in establishing the underlying
principle governing color confinement.

In U(1) lattice pure gauge 
theory  (no Higgs field), this same connection is seen to be
present, not in the defining variables,
but rather in the dual variables.   More specifically:
\begin{enumerate}
\item A field with non-zero magnetic monopole  charge, 
$\Phi$,
has been construc-ted\cite{fm}. It is a  composite  4-form
living on hypercubes constructed from
gauge fields.  There are also monopole current 3-forms.
On the dual lattice this monopole operator is a 0-form living
on dual sites.  The monopole currents are 1-forms 
living on dual links.  These currents either form closed loops
or end at monopole operators.   
The monopole operator has a non-zero vacuum expectation value 
in the dual superconducting
phase, $\langle \Phi \rangle \neq 0$, 
thereby signaling the spontaneous breaking 
of the U(1) gauge symmetry.
\item  Dual Abrikosov vortices have been seen 
in simulations\cite{shb,zfs}.  
They are identified
by the signature relationship between the electric field and the
curl of the monopole current in the transverse profile of the
vortex.  The  dual coherence length, $\xi_d$  measures the
characteristic distance from a dual-normal-superconducting
boundary over which the dual-superconductivity  turns on.
The dual London penetration length, $\lambda_d$ measures the attenuation
length of an external field penetrating the dual-superconductor.
 The dual photon mass  $ \sim 1/\lambda_d$ and
the dual Higgs mass $ \sim 1/\xi_d $.
\end{enumerate}

A signal  $\langle \Phi \rangle \neq 0$ without the consequent signal
of a dual photon mass does not imply confinement.  An observation
of a dual photon mass, i.e. vortex formation, 
without $\langle \Phi \rangle \neq 0$ does
not reveal the underlying principle governing the phenomenon.  

\subsection{Higgs effective theory}

The Higgs theory,  treated as an effective theory, i.e. limited 
to classical solutions,  and
considered in the dual sense, provides a model for
interpreting simulations of the pure gauge theory that can reveal these
important connections.  
Recalling the Higgs' current
\be
J_{\mu} = - \frac{i e}{2} \left( \phi^* 
\left(\partial_{\mu} - ie A_{\mu}\right) \phi -  
\phi
\left(\partial_{\mu} + ie A_{\mu}\right) \phi^* 
\right),
\label{bill}
\ee
and spontaneous gauge symmetry breaking through a constraint Higgs
potential
\be
\theta = v e^{i e \omega(x)}, \;\;\;\; v  = constant,
\label{bob}
\ee
leads to the London theory of a type II superconductor.
Using Eqns.(\ref{bill}) and (\ref{bob}) we obtain
\be
&&J_{\mu} = - e^2 v^2 \left( A_{\mu} - \partial_{\mu} \omega \right), 
\label{fred} \\ \nonumber \\
&&\left( \partial_{\mu} J_{\nu} -  \partial_{\nu} J_{\mu}\right)
+  m_{\gamma}^2 
\left( \partial_{\mu} J_{\nu} -  \partial_{\nu} J_{\mu}\right) = 0, 
\nonumber\\ \nonumber \\ 
&&{\bf \nabla} \times {\bf J} + \frac{1}{\lambda^2} {\bf B} = 0,
\nonumber
\ee
where 
\be
 e^2v^2 = m_{\gamma}^2 = \frac{1}{\lambda^2}.
 \label{vac}
\ee
Using Ampere's law $ \mb{\nabla}  \times \mb{B} = \mb{J} $, we obtain 
\ben
\mb{\nabla}^2 \mb{B} = \mb{B} /\lambda^2,
\nonumber
\een
identifying $\lambda$ as the London penetration depth.

If the manifold is multiply connected, then the gauge term in 
Eqn.(\ref{fred}) can contribute, as long as $e \omega(x)$ is
periodic, with period $2 \pi$ on paths that surround a hole.
\ben
\int_{S} (\mb{B}+ \lambda^2 \mb{\nabla} \times \mb{J} )\cdot \mb{n} d a &=&
\oint_{C} (\mb{A}+ \lambda^2 \mb{J} )\cdot d\mb{l},  \\ \\
&=&
\oint_{C}  \mb{\nabla}   \omega \cdot d\mb{l}, \\ \\ &=&
N \frac{2 \pi}{e} =N e_m = \Phi_{m}.
\nonumber
\een
where $N$  quanta of magnetic flux penetrates the hole in the manifold.
In real superconductors,  the hole is a consequence of the large 
magnetic field at the center which drives the material normal.

A cylindrically symmetric vortex solution is given by
\ben
\mb{B} + \lambda^2 \mb{\nabla} \times \mb{J} &=&
\Phi_{m} \delta^2(\mb{r} _{\perp}) {\bf n}_z,
\nonumber \\
\nonumber \\
  ( 1 - \lambda^2\mb{\nabla} _{\perp}^2 ) B_z(r_{\perp}) &=& 
\Phi_{m} \delta^2(\mb{r} _{\perp}), 
\nonumber \\
\nonumber \\
B_z(r_{\perp}) &=& 
\frac{\Phi_{m}}{2 \pi \lambda^2} K_{0}(r_{\perp}/\lambda).
\nonumber
\een
The delta function core of this vortex is normal, i.e. no spontaneous 
symmetry breaking, and the exponential
tail of the vortex is a penetration depth effect at the 
superconducting-normal boundary.  
The key point is that the modulus of the Higgs field must be independent
of position to get these idealized vortices.  For a 
``Mexican hat" Higgs potential, there is a coherence length setting the
length scale from a normal-superconducting boundary 
over which the vacuum expectation value of the Higgs field changes from 
zero to its asymptotic value.

On the lattice, the same phenomenon occurs.  We can
generate vortices  from finite configurations.  In the continuum
these objects are singular.   Since the lattice formulation 
is based on group elements, rather than the Lie algebra the 
 periodic behavior of the compact manifold is manifest. 
 This gives the 
 $2 \pi N$  ambiguity in the group angle 
 leading to N units of quantized flux.
 To see how this works,  consider the lattice Higgs action
\ben
S &=& \beta \sum_{x,\mu > \nu} (1 - cos\;\theta_{\mu\nu}(x)) 
\nonumber\\
 &&- 
    \kappa \sum_{x,\mu}(\phi^{*}(x)e^{i\theta_{\mu}(x)}  
                        \phi(x+\mu) + H.c.) +
    \sum_x V_{Higgs}(|\phi(x)|^2 ),
\label{e1}
\een
where 
$\theta_{\mu\nu}(x)$ is the curl of the gauge field,
\ben
    \theta_{\mu\nu}(x) = 
              \Delta^{+}_{\mu} \theta_{\nu}(x) -
                        \Delta^{+}_{\nu} \theta_{\mu}(x),
\een
and where $\phi(x)$ is the Higgs field and 
$\phi(x+\mu)$ refers to the Higgs field at the
neighboring site in the $\mu$ direction and $\Delta^{+}_{\nu}$ is the 
forward difference operator.  
The electric current is given by
\ben
 \frac{a^3}{e \kappa} J^{e}_{\mu}(x) =  
Im ( \phi^{*}(x)e^{i \theta_{\mu}(x)} 
     \phi(x+\mu)),
\label{e2}
\een
where $a$ is the lattice spacing.
Let us choose a Higgs potential that constrains the Higgs field 
$ |\phi(x)| = 1$. Then if
\ben
 \sin[\theta + 2N\pi] \approx \theta,
\label{e4.1}
\een
we find a relation between the field strength 
tensor and the curl of the current
\be
F_{\mu \nu} 
   - \frac{a^2}{e^2 \kappa}\frac{(  \Delta^+_{\mu} J^{e}_{\nu}(x) 
                            - \Delta^+_{\nu} J^{e}_{\mu}(x))}{a}
     = \frac{2\pi N}{e} \frac{1}{a^2} 
    =  N e_m \frac{1}{a^2},
\label{e5} 
\ee
where
\ben
ea^2 F_{\mu \nu}  = \sin[  \theta_{\mu}(x)
                         + \theta_{\nu}(x + \mu) 
                         - \theta_{\mu}(x + \nu)
                         - \theta_{\nu}(x)].
\label{e4} 
\een
If $N = 0$ then this is a London relation which implies a Meissner
effect.  For $N \neq 0$ there are
$N$ units of quantized flux penetrating that plaquette, indicating 
the presence of an Abrikosov vortex.

\subsection{Pure U(1) gauge theory}

In a pure U(1) lattice gauge simulation (without Higgs field),  lattice
averages over many configurations exhibit superconductivity in the dual
variables.  The superconducting current carriers are monopoles.  They
can be defined in a natural way on the lattice using the
 DeGrand Toussaint\cite{dt} construction.   
 They arise in non-singular configurations again because of
the $2 \pi N$ ambiguity in group elements.  These are the magnetic
charge carriers for dual superconductivity.
For a review of the monopole construction and vortex 
operators in U(1) gauge theory see e.g. the 
1995 Varenna Proceedings\cite{hay}.

As a brief summary, 
consider the unit 3-volume on the lattice at fixed $x_4$.   
The link angle is compact, 
$-\pi < \theta_{\mu} \leq \pi$.
The plaquette angle  is also compact, 
$-4\pi < \theta_{\mu\nu} \leq 4\pi$ and defined 
\ben
   ea^2 F_{\mu\nu}(x)= \theta_{\mu\nu}(x) = 
              \Delta^{+}_{\mu} \theta_{\nu}(x) -
                        \Delta^{+}_{\nu} \theta_{\mu}(x),
\een
where $a$ is the lattice spacing.
This  measures the electromagnetic flux through the face.   
Consider a configuration in which the absolute value of the
link angles,  $|\theta_{\mu}|$,  making
up the cube are all  small compared to $\pi/4$.  
 Gauss' theorem
applied to this cube then clearly gives zero total flux. 
Because of the $2\pi$ periodicity of the action we 
decompose the plaquette angle into two parts
\be
   \theta_{\mu\nu}(x) = \bar{\theta}_{\mu\nu}(x) + 2 \pi n_{\mu\nu}(x).
 \label{thetabar}
\ee
where $-\pi < \bar{\theta}_{\mu\nu} \leq \pi$.
If the four angles making up one of the six plaquette are adjusted 
so that e.g. $\theta_{\mu\nu}>\pi$ then
there is a discontinuous change in $\bar{\theta}_{\mu\nu}$  by $-2 \pi$
and a compensating change in $n_{\mu\nu}$.
  We can
clearly choose the configuration that leaves the plaquette angles on
all the other faces safely away from a discontinuity.  We then define 
a Dirac string $n_{\mu\nu}$ passing through this face (or better a 
Dirac sheet since the lattice is 4D).
 $\bar{\theta}_{\mu\nu}$ measures the electromagnetic flux
through the face.

This construction gives the following definition of the magnetic monopole current.
\be
\frac{a^3}{e_m}J^m_{\mu}(x)= \epsilon_{\mu \nu \sigma \tau}\Delta^+_{\nu}
\bar{\theta}_{\sigma \tau}(x). 
 \label{magcur} 
\ee
This lives on dual links on the dual lattice.   
Although Eqn.(\ref{thetabar}) is
not gauge invariant, Eqn.(\ref{magcur}) is.  Further it is a
conserved current, satisfying the 
conservation law $\Delta^+_{\mu}J^m_{\mu}(x) = 0$.

In simulations of a pure U(1) gauge theory we find that lattice
averages give a relation similar to Eqn.(\ref{e5}), but in the dual
variables
\ben
\la^*F_{\mu \nu}\ra
   - \lambda_{d}^2\frac{ \la \Delta^-_{\mu}  J^m_{\nu}(x) 
                            - \Delta^-_{\nu}  J^m_{\mu}(x) \ra }{a}
     = N  e  \frac{1}{a^2}, 
\label{e6} 
\een
where $^*F_{\mu \nu}$ is dual of $F_{\mu \nu}$.
This is the signal for the detection of dual vortices\cite{hay}.

\section{Non-Abelian theory}

The link of these considerations to confinement 
in  non-Abelian gauge theory  is through the
Abelian projection\cite{thooft1,klsw}.  One first fixes 
to a gauge while
preserving U(1) gauge invariance.   The non-Abelian
gauge fields can be parametrized in terms of a U(1) gauge field and charged
coset fields.  The working hypothesis is that operators 
constructed from the  U(1) gauge field alone, i.e. Abelian
plaquettes, Abelian Wilson loops, Abelian Polyakov lines 
and monopole currents,  will exhibit
the correct large distance correlations relevant for confinement.

In the continuum limit the maximal Abelian gauge condition is
\ben
\left( \partial_{\mu} \pm g A^{3}_{\mu}(x)\right)
 A^{\pm}_{\mu}(x) = 0.
\een
This is achieved on the lattice by a 
gauge configuration that maximizes ${ R}$, where
\ben
{ R}[U] &\equiv& \sum_{n,\mu} \frac{1}{2} 
Tr \left( \sigma_3 U_{\mu}(n) \sigma_3 U_{\mu}^{\dagger}(n )\right),
\een
and where $U_{\mu}(n)$ is the 
link starting a site $n$ and extending in
the $\mu$ direction.
\ben
U_\mu (n) &=& \left (
\begin{array}{cc}
\cos (\phi_\mu (n)) e^{{\textstyle i\theta_\mu (n)}} &
\sin (\phi_\mu (n)) e^{{\textstyle i\chi_\mu (n)}} \\& \\
-\sin (\phi_\mu (n)) e^{{\textstyle -i\chi_\mu (n)}} &
\cos (\phi_\mu (n)) e^{{\textstyle -i\theta_\mu (n)}} \\
\end{array}
\right). 
\nonumber
\een

After gauge fixing, the SU(2) link matrices may be decomposed in a 
`left coset' form:
\be
 U_\mu (n) = \left (
\begin{array}{cc}
\cos (\phi_\mu (n))  &
\sin (\phi_\mu (n)) e^{i\gamma_\mu (n)} \\& \\
-\sin (\phi_\mu (n)) e^{ -i\gamma_\mu (n)} &
\cos (\phi_\mu (n))  \\
\end{array}
\right )
\left (
\begin{array}{cc}
e^{ i\theta_\mu (n)} & 0 \\
\\
0 & e^{ -i\theta_\mu (n)} \\
\end{array}
\right ), 
\ee
Under a U(1) gauge transformation, 
$ \left\{ g(n) = \exp \left[ i \alpha(n) \sigma_3 \right] \right\} $,
\be
\theta_{\mu}(n) \rightarrow \theta_{\mu}(n) + \alpha(n) -
\alpha(n + \hat{\mu}) 
\mbox{\hspace{2em}}
\gamma_{\mu}(n) \rightarrow \gamma_{\mu}(n) + 2 \alpha(n)
\ee
In other words, the left coset field derived from the
link $U_\mu(n)$ is a doubly charged matter field living on the
site $n$ and is invariant under U(1) 
gauge transformations at neighbouring sites.  

The $c_\mu \equiv \cos (\phi_\mu)$ 
are real--valued fields which near the continuum $\sim 1
+ O(a^2)$ where $a$ is the lattice spacing. The off--diagonal 
$w_\mu \equiv \sin (\phi_\mu) e^{i\gamma_\mu}$
become the charged coset fields $g a W_\mu(x)$, and $\theta_\mu$ the
photon field $g a A^3_\mu(x)$. [The SU(2) coupling $\beta =
\frac{4}{g^2}$ in 3+1 dimensions.]

 The static potential constructed from Abelian links
gives as definitive a signal of confinement
as the gauge invariant static potential as found by 
Suzuki et.al.\cite{suzuki,ss}, Stack et.al.\cite{sw} and
Bali et. al.\cite{bbmps}.  Bali et.al. find the 
Abelian string  tension calculation gives $0.92(4)$ times the 
full string tension for $\beta = 2.5115$.  Whether this approaches
$1.0$ in the continuum limit remains to be seen.

Equation (\ref{vac}) gives the connection between the non-zero vacuum expectation
value of an order parameter and the photon mass or equivalently the
London penetration depth.   

This connection between order parameter and penetration depth is
the key to connecting spontaneous symmetry breaking to vortex
formation and hence confinement.  
Dual superconductivity  studies seek to establish a connection,
of course calculated from the original variables.

Di Giacomo et.al.\cite{giacomo,gp,baeg,glmp,ddpp} have
reported  extensive studies of the behavior of the
order parameter for the dual theory (denoted disorder parameter) at 
the confining transition.  See also Polikarpov et. al.\cite{cpv1}. Our group and others have reported
dual vortex formulation between sources allowing the determination
of the London penetration depth\cite{sbh,bdh,bali98}.  

We can demonstrate  a qualitative
correspondence between these two indicators of dual superconductiviy
 in that they are both
observed and both show the correct behavior on the two sides of the
transition.   However technical difficulties have eluded a direct
comparison check of the dual form of 
Eqn.(\ref{vac})

Figure 1 shows a plot of $\rho = 2 \frac{d}{d \beta} \ln \la \mu \ra$
as a function of $\beta$ near the transition. The spike indicates 
step behavior in $\la \mu \ra$ at the transition.

Electric dual vortices between sources
are well established\cite{sbh,bdh,bali98} in this gauge.  The typical behavior
is shown in Fig. 2.  The London relation is seen in the confining case
for transverse distances larger than about two lattice spacing. The 
dual coherence length $\xi_d \approx 2$, i.e. the onset of the violation
of the London relation.  The unconfined case is also shown where curl {\bf J}
is almost zero, and the electric field falls more gradually than in 
the confining case.

Bali et al\cite{bali98}
 have done a large scale simulation of these vortices.  Again
fitting the electric field and the monopole current 
to the Ginzburg-Landau theory their results are shown in Fig 3 and 4. 
Their data is very well described by the two G-L parameters
$\lambda = 1.84(24)$ and $\xi = 3.10(40)$ The ratio G-L parameter
$\kappa = \lambda/\xi = 0.59(14)$.  They find the total flux in 
the vortex, $\Phi =1.10(2)$ in units of the quantized flux.
They concluded tentatively that
type I dual superconductivity is indicated.

\section{Definition of electric field strength}

Central to finding the effective theory is the definition of
the field strength operator in the Abelian projected theory,
entering not only in the vortex profiles but also in the
formula for the monopole operator. All definitions should be
equivalent in the continuum limit, 
but use of the appropriate lattice
expression should lead to a minimization of discretization
errors.

In a recent paper\cite{dhh} we exploited 
lattice symmetries to derive such an
operator that satisfies Ehrenfest relations; Maxwell's 
equations for ensemble averages irrespective of
lattice artifacts.  This gives a precise lattice definition of current
and charge density independent of lattice size, and independent
of the continuum limit.

In the Abelian projection SU(2) link variables are parametrized
by U(1) links and  charged coset fields.
The latter  are normally discarded in Abelian
projection, as are the ghost fields arising from the 
gauge fixing procedure. Since the remainder of the SU(2)
infrared physics must arise from these, an understanding of
their role is central to completing the picture of full
SU(2) confinement.

In the maximal Abelian gauge the
supposedly unit charged Abelian Wilson loop has an
upward renormalization of charge due to this current. 
 A localised cloud of like polarity charge is
induced in the vacuum in the vicinity of a source, producing
an effect reminiscent of the antiscreening of charge in 
 QCD. In other gauges studied, the analogous current
is weaker, and acts to {\em screen} the source.

We show that this current can be quantitatively written as
a sum of terms from the coset and ghost fields. The contribution
of the ghost fields in the maximally Abelian gauge in
this context is found to be small. The effect of the
the Gribov ambiguity on these currents is argued to
be  slight.

\subsection{U(1)}

We first introduce and review the method due to Zach et al
\cite{zach95},
in pure U(1) theories 

Consider a shift in the U(1) link angles in the partition function 
containing a Wilson loop source term
\ben
Z_W(\{\theta^s_{\mu}\}) = \int [d (\theta_{\mu} + \theta^s_{\mu}) ] \;\;
\sin \theta_W  \;\;e^{\beta \sum \cos \theta_{\mu \nu}}.
\een
Since the Haar measure is invariant under this shift, $Z_W$ is constant
in these variables.  Absorbing the shift into the integration variable
and the taking the derivative
\ben
\frac{\partial}{\partial \theta^s_{\nu}(x_0)} Z_W = 0,
\een
we get the relation
\ben
0 = \int [d \theta] \;
(\cos \theta_W - \sin \theta_W \;\beta \Delta_{\mu} \sin \theta_{\mu \nu})
\;e^{\beta \sum \cos \theta_{\mu \nu}}.
\een
This can be cast into the form
\ben
\la \Delta_{\mu} F_{\mu \nu} \ra_W  \equiv 
\frac{\la \sin \theta_W \;\;
\Delta_{\mu}\frac{1}{e} \sin \theta_{\mu \nu} \ra }
{\la \cos \theta_W \ra} =  e \delta_{x, x_W} = J_{\nu}.
\een

This is a well known technique to generate exact relations between
Green's functions that is useful in generating Ward identities, 
or Schwinger-Dyson equations, or in  this case we denote as
Maxwell-Ehrenfest relations.   We use the term Ehrenfest because 
it is the expectation value of what is normally a classical extremum
of the path integral - an Euler Lagrange equation.

\subsection{SU(2) no gauge fixing}

Before addressing the full problem we first generalize from
U(1) to SU(2) without the complication of gauge fixing.
\ben
Z_W(\{U^s\}) &=& \int [d (UU^s)]\;\; W_3(U)\;\; e^{\beta S(U)},
\een
The size of the source is irrelevant so we choose 
it to be the simplest case, i.e. a plaquette:
\ben
W_3 \equiv \frac{1}{2}Tr (U^{\dagger}U^{\dagger} U U i \sigma_3).
\een
We choose the shift to be in the $3$ direction
\ben
U^s_{\mu}(x_0) = \left(1 - \frac{i}{2}\epsilon_{\mu}(x_0) \sigma_3 \right).
\een
The invariance 
\ben
\frac{d}{d \epsilon_{\mu}(x_0)}Z_W = 0
\een
gives the Ehrenfest relation
\ben
\beta \frac{\la W_3 S_{\mu}  \ra}{\la W  \ra} =
 \delta_{x, x_W}.
\een

For $\beta=2.5$,   $\beta \la W_3 S_{\mu} \ra  = 0.0815(2)$,   and  
$\la W  \ra  = 0.0818(1)$, and the difference $= 0.0003(3)$.

The notation $S_{\mu}$ denotes an $\epsilon$ derivative\cite{dhh}.  
The denominator is just the ordinary plaquette.
\ben
W \equiv \frac{1}{2}Tr (U^{\dagger}U^{\dagger} U U )
\een

To cast this into the form of Maxwell's equation we
decompose the link into diagonal $D_{\mu}$ and off-diagonal parts
$O_{\mu}$
\ben
U_{\mu}(x) = D_{\mu}(x) + O_{\mu}(x)
\een
Further we simplify notation 
\ben
 \la  \cdots  \ra_W \equiv \frac{\la W_3 \cdots\ra }{\la W \ra}
\een

We then group terms involving the diagonal part on the left and 
take all terms having at least one factor of the off-diagonal
link to the right.
\ben
\left[\beta \la S_{\mu}  \ra_W \right]_{U = D} =  
\frac{1}{e} \la  \Delta_{\mu} F_{\mu \nu} \ra_W . 
\een
Finally, note
\ben
\delta_{x, x_W} = \frac{1}{e} J_{\nu}^{static} ,
\een
giving the final form of the Ehrenfest relation
\ben
\la \Delta_{\mu} F_{\mu \nu} \ra_W 
= \la J_{\nu}^{dyn.} \ra_W  + J_{\nu}^{static} .
\een
This then tells us how to choose a lattice definition of field 
strength that satisfies an Ehrenfest relation:
\ben
F_{\mu \nu} = 
\frac{1}{e} \frac{1}{2}
Tr(D^{\dagger}D^{\dagger} D D i \sigma_3)_{\mu \nu}
\een

\subsection{Gauge fixed SU(2), U(1) preserved}

The effect of gauge fixing gives
\ben
Z_W(\{U^s\}) = \int [d (UU^s)]\; W_3(U)\;\Delta_{FP} \; 
\delta[F]\; e^{\beta S(U)},
\een
where we have introduced
\ben
1 = \Delta_{FP}  \int \prod_{j,y} d g_j(y)  \prod_{i,x}
 \delta[F_i(U^{\{g_j(y)\}}; x)],
\een
and integrated out the $g$ variables in the standard way.

In this case  $Z_W$  is {\em not invariant}.  The shift 
is inconsistent with the gauge condition.   
However,  it is invariant under an infinitessimal shift together 
with an infinitessimal `corrective' gauge transformation
that restores the gauge fixing
\ben
G(x) = \left(1 - \frac{i}{2}\mb{\eta} (x) \cdot  \mb{\sigma} \right).
\een

Use of the invariance of the measure under combination of
a shift and a `corrective' gauge transformation we obtain
\ben
\left[ \frac{\partial}{\partial \epsilon_{\mu}(z_0)} + \sum_{k,z} \;\;
\frac{\partial \eta_k(z)}{\partial \epsilon_{\mu}(z_0)}
\frac{\partial }{\partial \eta_k(z)}
\right]Z_W = 0.
\een

In shorthand notation\footnote{See Ref.\cite{dhh}},
the Ehrenfest relation reads
\be
\lla (W_3)_{\mu}\biggr|_s + 
(W_3)_{\mu}\biggr|_g + W_3 \left(
\frac{(\Delta_{FP})_{\mu}}{\Delta_{FP}}\biggr|_s + 
\frac{(\Delta_{FP})_{\mu}}{\Delta_{FP}}\biggr|_g  +
\beta S_{\mu}\right)
\rra  = 0.
\label{terms}
\ee

Gauge fixing has introduced three new terms:
\begin{itemize}
\item $(W_3)_{\mu}\biggr|_g$ comes from the corrective gauge 
transformation acting on the source which is U(1) invariant but
not SU(2) invariant. \\
\item $\frac{(\Delta_{FP})_{\mu}}{\Delta_{FP}}\biggr|_s$ is due to
the shift of the Faddeev-Popov determinant.
\item $\frac{(\Delta_{FP})_{\mu}}{\Delta_{FP}}\biggr|_g$ is due to
the corrective gauge transformation of the Faddeev-Popov determinant.
\end{itemize}

The latter two derivatives are subtle. The key is to first consider
the constraint up to first order in the shift and the corrective
gauge transformations
\ben
F_i(x) + \frac{\partial F_i(x)}{\partial \epsilon(z_0)} d \epsilon(z_0)
+ \sum_{k,z} \frac{\partial F_i(x)}{\partial \eta_k(z)} d \eta_k(z) = 0,
\een
and then define the Faddeev-Popov matrix  as a derivative
of the corrected constraint.  
\ben
M_{i x; j y} + \delta M_{i x; j y}  = 
\frac{\partial}{\partial \eta_j(y)}\left\{
F_i(x) + \frac{\partial F_i(x)}{\partial \epsilon(z_0)} d \epsilon(z_0)
+ \sum_{k,z} \frac{\partial F_i(x)}{\partial \eta_k(z)} d \eta_k(z)
\right\} 
\een
Finally we evaluate the derivative using
\ben
\frac{\Delta_{\mu}}{\Delta}  = Tr[ M^{-1} M_{\mu}]
\een

A check of the Ehrenfest theorem is given in Table 1.  
Some of the
individual terms on the right hand side require a 
$2 N \times 2 N$ matrix 
inversion, wherer N is the lattice
volume.  Hence to test the result 
numerically, we chose as small a lattice as possible. The result
does not involve the size of the lattice which is $4^4$ in table 1.
The last column employees a different source.  The links making
up the plaquette are replaced by the diagonal parts only.

By separating the links into diagonal and off-diagonal
parts we get the final form of the Ehrenfest-Maxwell  relation.

\ben
\lla \Delta_{\mu} F_{\mu \nu} \rra = \lla J_{\nu}^{dyn.} \rra
+ J_{\nu}^{static}\biggr|_{s}
+ J_{\nu}^{static}\biggr|_{g}
+ \lla J_{\nu}^{FP}\biggr|_{s}\rra
+ \lla J_{\nu}^{FP}\biggr|_{g}\rra
 \een

The first term in the current comes from the excitation of the
charged coset fields,  the static term has an extra non-local
contribution coming from the corrective gauge transformation, 
and the last two contributions are from the ghost fields. 

These terms give a non vanishing charge density cloud around

a static source.  The lefthand side can be used as a lattice

operator to measure the total charge density.

\subsection{Abelian point charge has cloud of like charge}

As a simple application we use this definition of flux to
calculate div E on a source and the total flux away from the
source.  In Table 2,  we  see that the integrated flux on a plane
between the charges plus the integrated flux on a back plane of the torus
is larger than the div E on the source.  The interpretation 
is that  the bare charge is dressed with same polarity 
 charge by the interactions and the neighborhood has a cloud
 of like charge. Hence there is anti-screening.   This charge
 density has contributions from all terms in the Ehrenfest relation.

\subsection{Summary}
\label{sec_summ}

We have exploited symmetries of the lattice partition
function to derive a set of exact, non--Abelian identities which
define the Abelian field strength operator and a conserved
electric current arising from the coset
fields traditionally discarded in Abelian projection. The current has
contributions from the action, the gauge fixing condition and the
Faddeev--Popov operator. Numerical studies on small
lattices verified the identity to within errors of a few per cent. 
We have found the Faddeev--Popov current in particular to be unusually 
sensitive to
systematic effects such as low numerical precision and poor random
number generators, but the origin of any remaining, subtle biases,
if they exist, is not clear; we have
already considered all terms in the partition function.

In a pure U(1) theory the static quark potential may be measured
using Wilson loops that correspond to unit charges moving in
closed loops, as demonstrated by 
$|\langle \diffn{\nu} F_{\nu \mu} \rangle| = \delta_W$.
In Abelian projected SU(2) the same measurements in the maximally
Abelian gauge yield an asymptotic area law decay and a string
tension that is only slightly less than the full non--Abelian
value. In other gauges it is not clear that an area law exists ---
certainly it is more troublesome to identify.

We have seen that in the context of the full theory the Abelian
Wilson loop must be reinterpreted. The coset fields renormalize 
the charge of the loop as measured by 
$|\langle \diffn{\nu} F_{\nu \mu} \rangle|$
and charge is also induced in the surrounding vacuum. Full
SU(2) has antiscreening/asymptotic freedom of colour
charge, and in the maximally Abelian gauge alone have we seen
analogous behavior, in that the source charge is increased 
and induces charge of like polarity in the neighbouring vacuum.
Whether this renormalisation of charge can account for the
reduction of the string tension upon Abelian projection in 
this gauge is not clear. In other gauges, where Abelian
dominance of the string tension is not seen, the coset fields
appear to have a qualitatively different behavior, acting
to suppress and screen the source charge.

In conclusion,
the improved field strength expression defined by the Ehrenfest 
identity does not coincide with the lattice version of
\cite{bdh}
of 't Hooft's proposed field strength operator
\cite{thooft2}.
The Abelian and monopole dominance of the string tension invites a
dual superconductor hypothesis for confinement. If this is to be
demonstrated quantitatively such as by verification of a (dual) London
equation then a a careful understanding of the field strength operator
is required. The Ehrenfest identities may provide this
\cite{hartprog}.

\section{Acknowledgments}

I wish to thank  G. Di Cecio, A. Di Giacomo, J. Greensite, 
G. Gubarev, A. Hart,  M.I. Polikarpov Y. Sasai,  E. T.  Tomboulis 
and J. Wosiek for 
discussions.  This work was supported in part by United States 
Department of Energy grant DE-FG05-91 ER 40617.

\section{Figure captions}

Fig. 1. (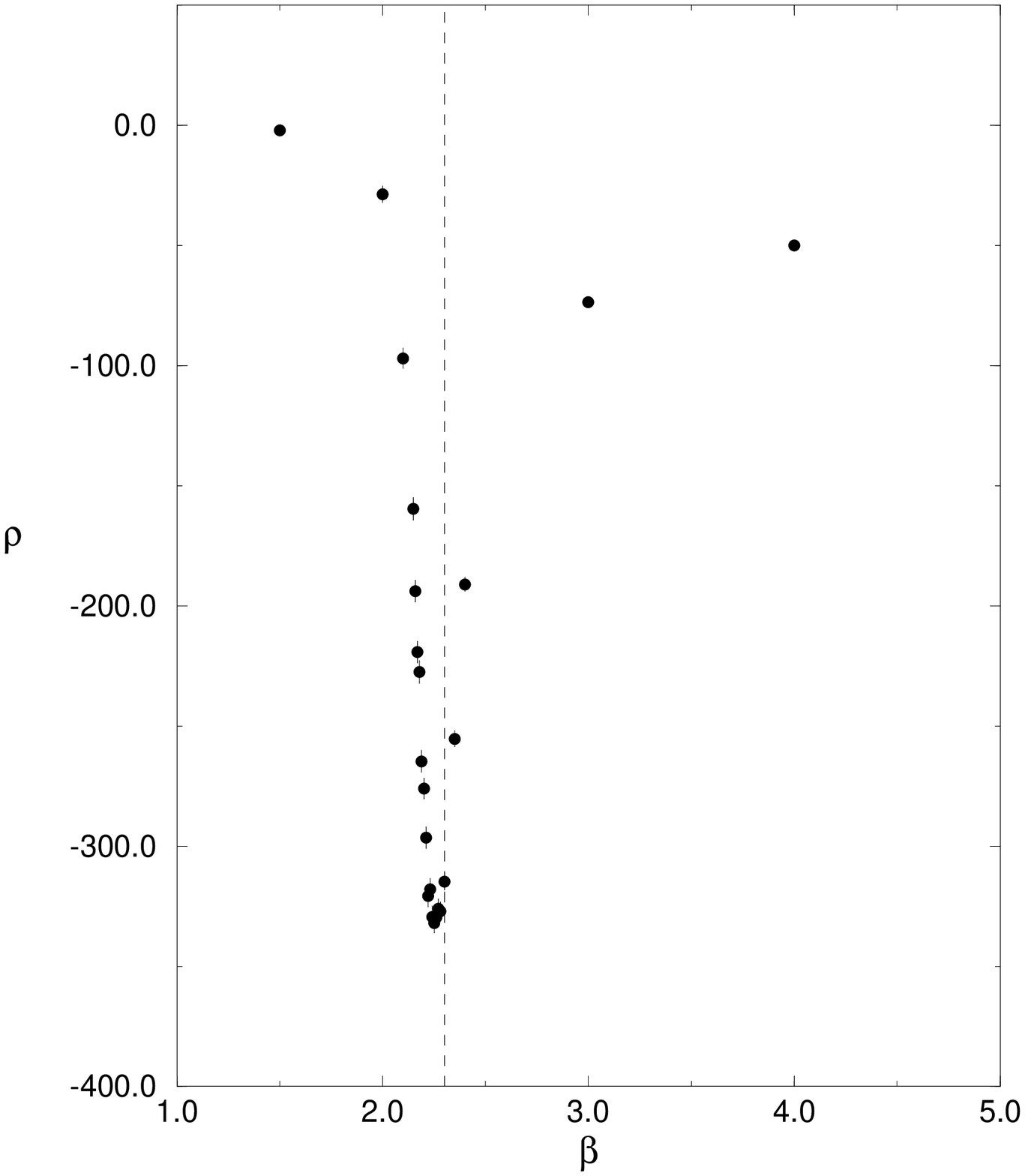)  (From ref. \cite{giacomo})
 $\rho$ vs. $\beta$ for SU(2) gauge theory.  The peak signals deconfining
 phase transition.  here monopoles are defined by the Abelian
 projection on Polyakov line.

Fig. 2.  (fig7.tex needs fig7 1.eps, fig7 2.eps and epsfig.sty)
(From ref. \cite{bdh})
Transverse profile of the electric field and curl of the
monopole current in the mid plane between a static $q \bar{q}$ pair 
in the maximal Abelian gauge at 
finite temperature for a confining (left) 
and unconfining (right) phase.

Fig 3.  (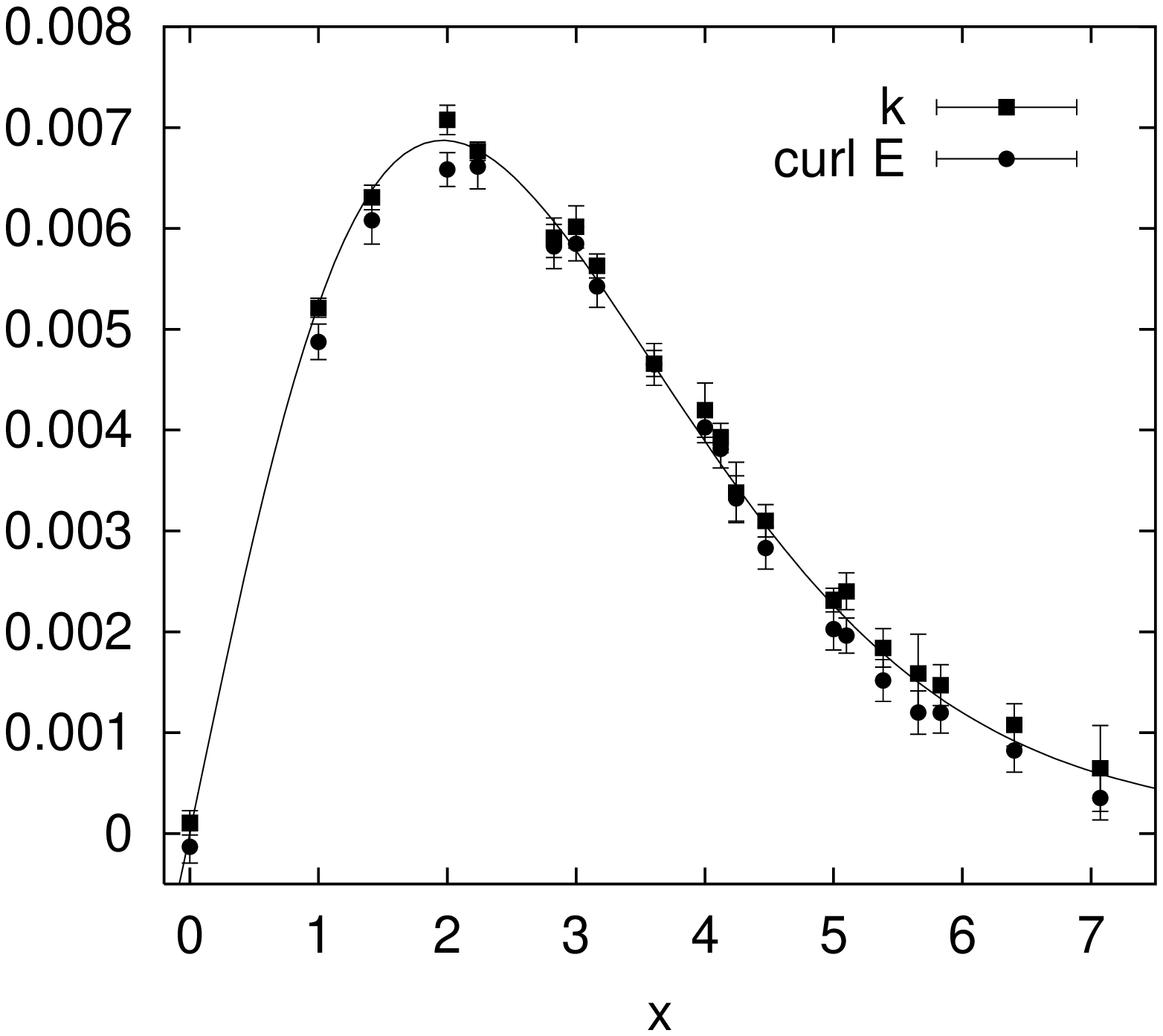)  (From \cite{bali98})
Check of the dual Ampere law in the dual vortex profile. 
 $\mb{E} $ is the electric field,  $\mb{k} $ is the monopole
 current.

Fig 4.  (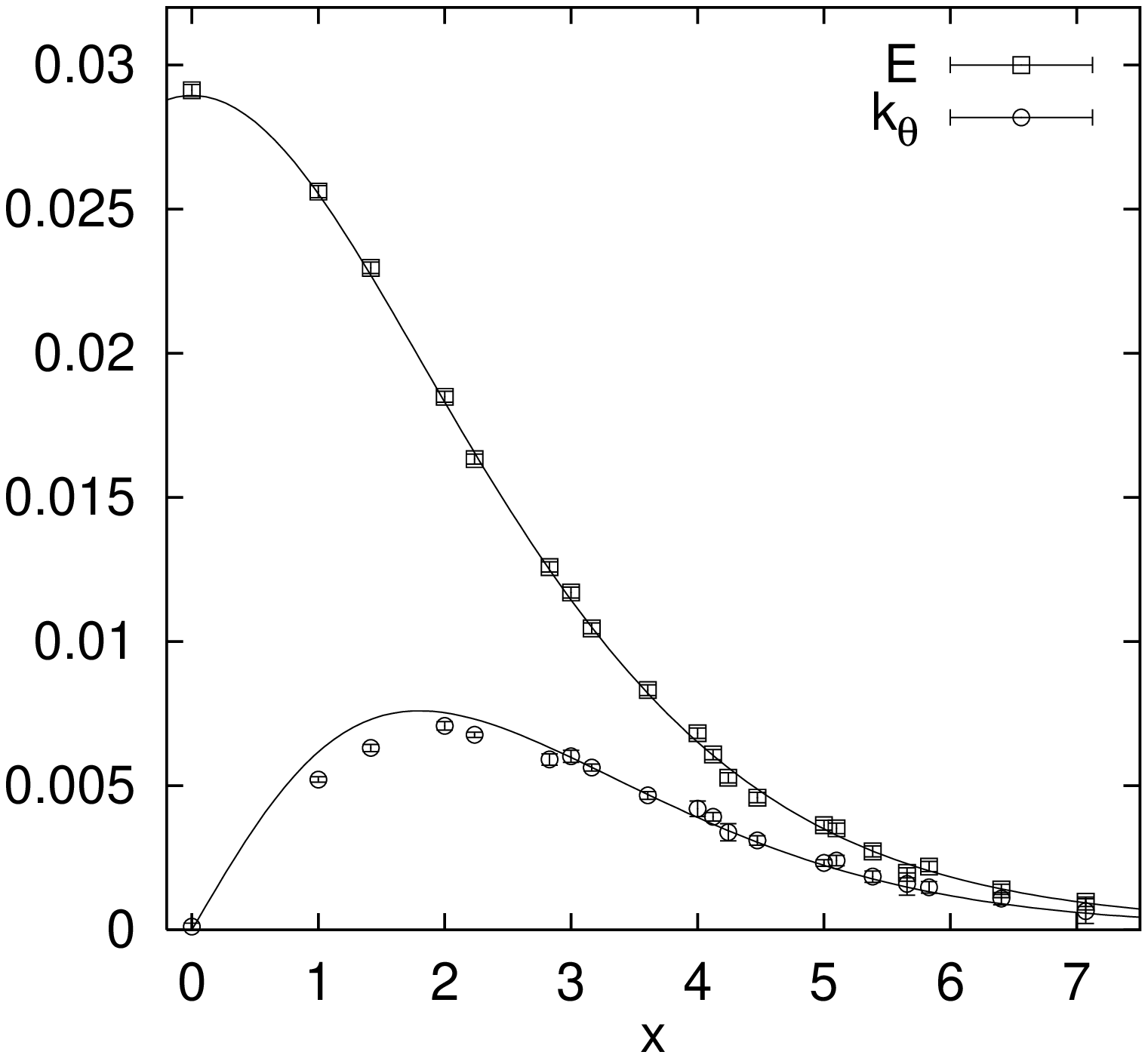)  (From \cite{bali98})
Fit of the electric field vortex profile $\mb{E} $ and the tangential 
component of the magnetic monopole current $\mb{k} _{\theta}$ to 
the Ginzburg Landau theory.

\newpage

\begin{table}
\begin{tabular}{lll} \hline
Source:\hspace{4cm} & $W_3$  \hspace{1cm}& $W_3(U \rightarrow D)$ \\ 
Ehrenfest term & &  \\  \hline
$\lla (W_3)_{\mu}\biggr|_s 
\rra $

 &  0.65468(10) \hspace{1cm}&  0.63069(20) \\ \\
 
 $\lla (W_3)_{\mu}\biggr|_g 
\rra$

 &  0.06095(7) \hspace{1cm}&  0.04463(4) \\ \\

$\lla W_3  \frac{(\Delta_{FP})_{\mu}}{\Delta_{FP}}\biggr|_g \rra
$
  
 &  0.00127(21) \hspace{1cm}&  0.00132(50) \\ \\
 
$\lla W_3  \frac{(\Delta_{FP})_{\mu}}{\Delta_{FP}}\biggr|_s \rra
$
  
 &  0.00529(3) \hspace{1cm}&  0.00564(3)  \\ \\
 
$\lla \beta (S)_{\mu}\biggr|_s 
\rra $

 &  -0.72246(68) \hspace{1cm}&  -0.68275(50)   \\ \\ \hline

Zero  & -0.00026(77)& -0.00045(64)

 \end{tabular}
\caption{Terms in the Ehrenfest relation, Eqn.(terms).  
The column labeled '$W_3$' corresponds to the source described in the text.
In the second column the links are replaced by the diagonal arts
of the links in order to test a second example. The theorem gives
zero for the sum. $\beta = 2.5$, $4^4$ lattice}
  
\end{table}

\begin{table}
\begin{tabular}{cccc} \hline
$\beta$ & $div E$(cl.pt.charge) 
 & $div E$ (on source) & total flux  
\\ 
& $= \frac{1}{\beta}$&& \\ \hline \\
10.0 &  0.1 & 0.1042(1) &  0.0910(8) (mid) \\ 
(almost&&& 0.0148(8) (back)\\ 
classical)&&& 0.1092(8) (total)\\

  \\ \hline \\
2.4 &  0.4166 & 0.5385(19) &  0.7455(70) (mid) \\ 
&&& 0.0359(72) (back)\\ 
&&& 0.7815(95) (total)\\ \\ \hline
\end{tabular}
\caption{$div E$ normalized to $\frac{1}{\beta}$ for a classical
point charge. Source is a  $3 \time 3$ Wilson loop.  $div E$ 
measured on a source.  Electric Flux measured on the midplane centered
on the Wilson loop.  Also included is the flux through a plane on
the far side of the torus, and the sum being the total flux.  $8^4$
lattice, $3 \times 3$ Wilson loop.}
\end{table}

\end{document}